\documentstyle[aas2pp4]{article}

\def\etal{{\it et al.}}
\def\comment#1{{}}

\begin{document}

\title{Identification of the Periodic Hard X-Ray Transient
GRO~J1849--03 with the X-Ray Pulsar GS~1843-02 = X1845-024 - a New
Be/X-Ray Binary}

\authoremail{kaaret@astro.columbia.edu}

\author{P. Soffitta\altaffilmark{1}, J.A. Tomsick\altaffilmark{2}, 
B.A. Harmon\altaffilmark{3}, E. Costa\altaffilmark{1}, E.C.
Ford\altaffilmark{2}, M. Tavani\altaffilmark{2}, S.N.
Zhang\altaffilmark{4,3}, and P. Kaaret\altaffilmark{2}}

\altaffiltext{1}{Istituto di Astrofisica Spaziale, CNR, C.P. 67,
I-00044 Frascati, Italy} 

\altaffiltext{2}{Columbia Astrophysics Laboratory, Columbia University,
550 W. 120th Street, New York, NY 10027}

\altaffiltext{3}{NASA/Marshall Space Flight Center, ES 84, Huntsville,
AL 35812}

\altaffiltext{4}{University Space Research Association/MSFC, ES 84,
Huntsville, AL 35812}

\begin{abstract}

We identify the periodic transient hard X-ray source GRO~J1849--03 with
the transient x-ray pulsar GS~1843-02 = X1845--024 based on the
detection of x-ray outbursts from X1845-024 coincident with hard x-ray
outbursts of GRO~J1849--03.  Based on its spin period of 94.8~s and its
orbital period of 241 days, we classify the system as a Be/X-ray
binary.

\end{abstract}


\section{Identification of GRO~J1849-03}

GRO~J1849-03 is a periodic transient hard x-ray source in the direction
of the 5-kpc arm (Zhang et al. 1996).  The source was discovered using
the Burst And Transient Source Experiment (BATSE) on the Compton
Gamma-Ray Observatory and shows recurrent hard x-ray  (20--100~keV)
outbursts with a period of $241 \pm 1$~days (Zhang et al. 1996).  The
source produced a hard x-ray outburst detected with BATSE in September
1996 lasting from MJD 50340 to MJD 50358 (Barret et al. 1996).  The
maximum of the outburst occurred near MJD 50347.

In an effort to determine an accurate position for GRO~J1849-03, an
observation was made with the Wide Field Camera (WFC) on the {\it
Satellite Italiano per Astronomia X} (BeppoSAX) on 17-18 September 1996
(MJD 50343-50344).  The observation was scheduled to occur during the
predicted hard x-ray outburst.  An integration of 7712~s was obtained
in two spacecraft orbits.  The data were analyzed using the standard
WFC software (version 101-104) and sources were extracted using the
Iterative Removal of Sources (IROS) software (Hammerslay 1986; in't
Zand 1992).  We chose to use a relatively hard band (5--21~keV) for the
analysis since the x-ray sources in the 5-kpc arm have high column
densities (Koyama et al. 1990a) and also since the BATSE detection of
the source extends down only to 20~keV.  No sources were detected
within the $1 \sigma$ BATSE error box above a flux of $1.4 \times
10^{-10} \, \rm erg \, cm^{-2} \, s^{-1}$ for the 5-21~keV band at $3
\sigma$ confidence.  Eleven sources were detected at high confidence in
the WFC field of view.  The only source detected near GRO~J1849-03 is
X1845-024 (Doxsey et al. 1977).  The WFC source is within 3' of the
known position of X1845-024; this is consistent with the position
uncertainty from the WFC.  The source flux is $(2.1 \pm 0.5 ) \times
10^{-10} \, \rm erg \, cm^{-2} \, s^{-1}$ for the 5-21~keV band. 
X1845-024 lies within $2 \sigma$ of the BATSE position for GRO~J1849-03
(Zhang et al. 1996) as shown in Fig. 1.

We produced a light curve for X1845-024, shown in Fig.~2, using the
All-Sky Monitor (ASM) on the {\it Rossi X-ray Timing Explorer} (RXTE)
to search for transient behavior similar to that observed from
GRO~J1849-03.  The light curve was produced from the ``definitive data
products'' provided by the ASM/RXTE teams at MIT and at the RXTE SOF
and GOF at NASA's GSFC.  Each point represents the ASM counting rate in
the 5--12~keV band averaged over 4 days.  Points with errors larger
than 0.3~c/s have been suppressed for clarity.  Inspection of the light
curve (Fig.~2) shows two outburst events.  The first starts on MJD
50342 and lasts until approximately MJD 50366.  The peak occurs near
MJD 50350-50354.  This x-ray outburst coincides with the September 1996
outburst of GRO~J1849-03 detected by BATSE.  The main parts of the
X1845-024 and the GRO~J1849-03 outbursts overlap.  A second outburst in
the x-ray light curve of X1845-024 occured during the interval MJD
50589-50598 with the peak near MJD 50591-50594.  Hard x-ray emission
was detected with BATSE from the direction of X1845-024 over the
interval MJD 50585-50597.  The second outburst found in the ASM is
approximately 240-243 days after the first.  The period between
outbursts matches the known period of GRO~J1849-03 (Zhang et al. 1996).

Our WFC observation shows that X1845-024 is the only bright source
within the $2 \sigma$ error box for GRO J1849-03 near the time of the
September outburst.  The detection of two x-ray outbursts from
X1845-024 simultaneous with hard x-ray outbursts detected by BATSE from
GRO~J1849-03 lead us to conclude that GRO J1849-03 and X1845-024 are
the same object.  The outburst of GS~1843-02 detected by Ginga (Koyama
et al. 1990b) occurred at a time only $1 \sigma$ away from the
extrapolation of the BATSE ephemeris for GRO~J1849-03 (Zhang et al.
1996).  It is likely that GS~1843-02 is also X1845-024 (Makino et al.
1988).

The composite BATSE/ASM spectrum for the interval MJD 50331.5-50358.5
is adequately fit by a exponentially cutoff power law with a photon
index of $-1.4 \pm 0.4$, a cutoff energy of $42 \pm 25 \, {\rm keV}$,
and a column density of $2.5 \pm 1.0 \times 10^{23} \, {\rm cm}^{-2}$.
These are consistent with the spectral parameters and column density
found for GS~1843-02 (Koyama et al. 1990b).  This lends support to the
identification of GS~1843-02 and X1845-024.

We searched for pulsations with periods near 94.8~s in the WFC data. 
However, the results were inconclusive due to the weakness of the
source and the presence of the bright and highly variable source
GRS~1915+105 in the WFC field of view.  We did not search for
pulsations in the ASM data as the 94.8~s period is nearly commensurate
with the 90~s duration of individual ASM dwells.  Pulsations at the
94.8~s period of X1845-024 have not been detected from GRO~J1849-03
except for a marginal detection during the latter part of one outburst
(Zhang et al. 1996).  However, the upper limit of 35\% pulsed fraction
in the 20--50~keV band is not strongly inconsistent with the 40\%
pulsed fraction observed by Ginga at lower energies (Koyama et al.
1990b).

\section{The nature of X1845-024}

GS~1843-02 is an x-ray pulsar with a pulse period of 94.8~s (Koyama et
al. 1990a).  Given the association of GRO~J1849-03, GS~1843-02, and
X1845-024, we can classify the system based on its pulse and orbital
period via the method of Corbet (1986).  We assume that the 241~day
outburst recurrence time is the orbital period.  Fig.~3 shows a plot of
pulse versus orbital period for high-mass x-ray binaries, a `Corbet
diagram', indicating the position of X1845-024.  The object clearly
lies in the region of the diagram occupied by Be-star systems. The
location of the source, coincident with the 5 kpc arm, lends some
support to the Be-star binary identification, as IR observations of the
region show that it contains a population of young stars (Hayakawa et
al. 1981) with ages consistent with those estimated for Be/X-ray
binaries (van den Heuvel \& Rappaport 1987).  If the source lies within
the 5-kpc arm, at a distance of 10~kpc, the unabsorbed luminosity of
the source during outburst is $6 \times 10^{36} \, \rm erg \, s^{-1}$. 
This is consistent with luminosities observed from long orbital period
Be/X-ray binaries.

X1845-024 lies slightly below the main track of Be systems on the
Corbet diagram.  The observed correlation between spin and orbital
periods of x-ray pulsars in Be-star binaries is believed to depend on
the interaction between the neutron star magnetic field and the Be-star
outflow.  X1845-024 may have a relatively small value of the density
vs. radius index ($n$ between 3  and 3.25) and/or a low initial outflow
velocity ($v_0 \leq 5 \, \rm km \, s^{-1}$) compared to other Be
systems (Waters \& van Kerkwijk 1989).  Since no centrifugal barrier is
expected for a neutron star with a spin period of 94.8~s, the duration
of x-ray outbursts should be directly related to the size or width of
the Be-star equatorial outflow.  Identification of the optical or IR
companion of X1845-024 and a study of its mass outflow may help provide
a better understanding of the spin and orbital period correlation.

\newpage


\begin{figure}[ht] \epsscale{1.0} \plotone{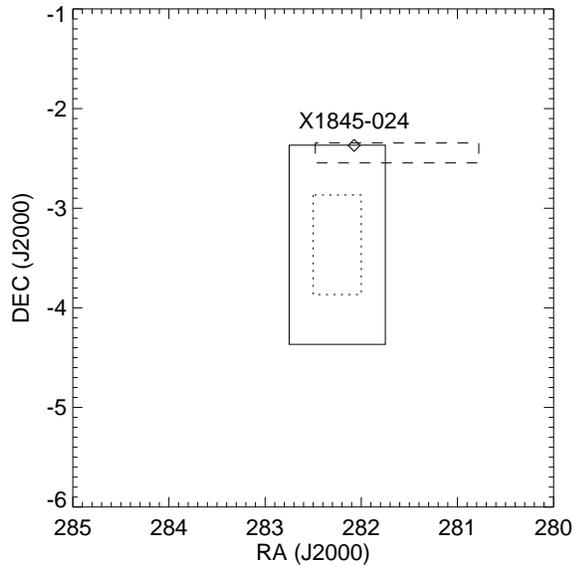}
\figcaption{Sources near GRO~J1849-03.  The only source in this region
detected with the WFC is X1845-024, shown as a diamond.  The dotted
square is the $1\sigma$ confidence error box for GRO~J1849-03 and the
solid square is the $2\sigma$ confidence error box (Zhang et al.
1996).  The dashed square is the error box for GS~1843-02 (Koyama et
al. 1990b).} \label{fig_wfcmap} \end{figure}

\begin{figure}[ht] \epsscale{1.0} \plotone{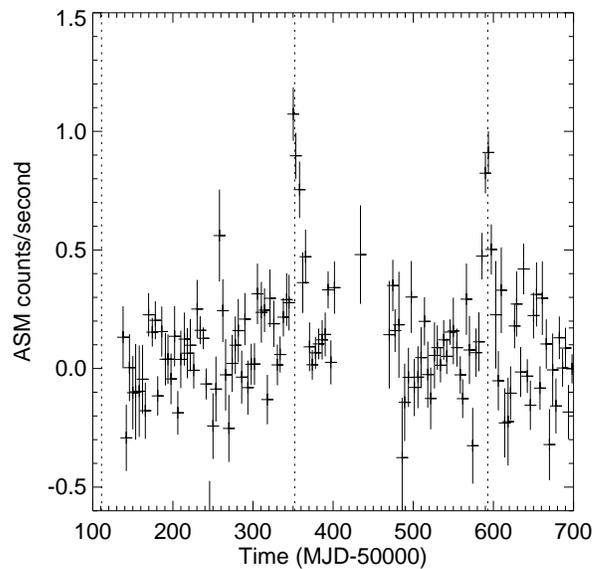}
\figcaption{ The x-ray (5--12~keV) light curve for X1845-024 from the
ASM/RXTE.  Each point represents an average over 4 days and points with
errors larger than 0.3~c/s have been suppressed.  The dotted lines show
241 day intervals and are drawn at MJD 50111, 50352, and 50593.}
\label{fig_asmlc} \end{figure}

\begin{figure}[ht] \epsscale{1.0} \plotone{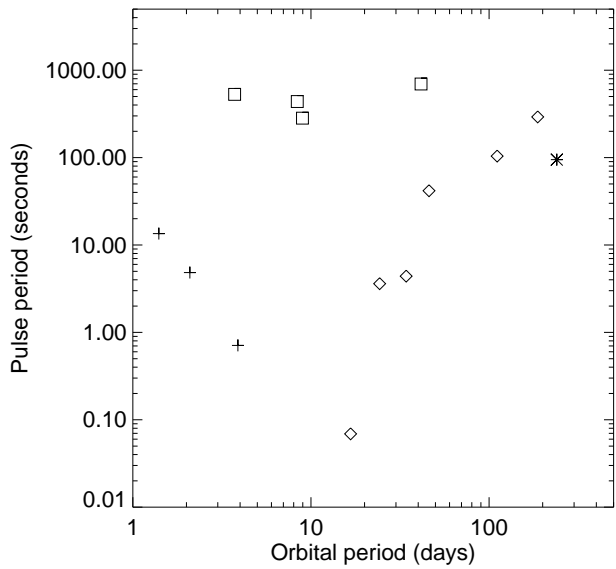}
\figcaption{Spin period versus orbital period for various high mass
x-ray pulsars (following Corbet 1986).  Diamonds indicate Be primaries,
boxes indicate supergiant primaries, crosses indicate likely Roche lobe
filling primaries, and the asterisk is for X1845-024.}
\label{fig_corbet} \end{figure}

\end{document}